\documentclass[prd,aps,nofootinbib,showpacs,showkeys]{revtex4}
\def\Xint#1{\mathchoice
   {\XXint\displaystyle\textstyle{#1}}%
   {\XXint\textstyle\scriptstyle{#1}}%
   {\XXint\scriptstyle\scriptscriptstyle{#1}}%
   {\XXint\scriptscriptstyle\scriptscriptstyle{#1}}%
   \!\int}
\def\XXint#1#2#3{{\setbox0=\hbox{$#1{#2#3}{\int}$}
     \vcenter{\hbox{$#2#3$}}\kern-.5\wd0}}

\def\dashint{\Xint-}
\usepackage{graphicx}
\usepackage{bm}
\usepackage{amssymb}
\usepackage{amsmath}
\usepackage{euscript}

\begin{document}

\title{Quantum Fluctuations of a Coulomb potential}

\author{Kirill~A.~Kazakov\thanks{E-mail address: $kirill@theor.phys.msu.su$}}

\affiliation{Department of Theoretical Physics,
Physics Faculty,\\
Moscow State University, $119899$, Moscow, Russian Federation}

\begin{abstract}
Long-range properties of the two-point correlation function of the
electromagnetic field produced by an elementary particle are
investigated. Using the Schwinger-Keldysh formalism it is shown
that this function is finite in the coincidence limit outside the
region of particle localization. In this limit, the leading term
in the long-range expansion of the correlation function is
calculated explicitly, and its gauge independence is proved. The
leading contribution turns out to be of zero order in the Planck
constant, and the relative value of the root mean square
fluctuation of the Coulomb potential is found to be $1/\sqrt{2},$
confirming the result obtained previously within the S-matrix
approach. It is shown also that in the case of a macroscopic body,
the $\hbar^0$ part of the correlation function is suppressed by a
factor $1/N\,,$ where $N$ is the number of particles in the body.
Relation of the obtained results to the problem of measurability
of the electromagnetic field is mentioned.
\end{abstract}

\maketitle

\section{Introduction}

Fluctuations in the values of physical quantities is an
inalienable trait of all quantum phenomena, which finds its
reflection in the probabilistic nature of the quantum-mechanical
description of elementary particle kinematics as well as of the
field-theoretic treatment of the fundamental interactions.
Investigation of fluctuations constitutes an essential part in
establishing quasi-classical conditions, and therefore in dealing
with the issue of correspondence between given quantum theory and
its classical original.

Precise identification of the quasi-classical conditions is
especially important in investigation of the measurement process.
In this process, an essential role is played by the measuring
device ``recording'' the results of an observation, and the
question of primary importance is to what extent this device can
be considered classically. Furthermore, investigation of
fluctuations in the observable quantities themselves is a
significant ingredient in the theoretical treatment of the
measurement process, in particular, in verifying conformity of the
formalism of quantum theory with the principal realizability of
measurements.

In the latter aspect, the issue of quantum fluctuations of the
electromagnetic field was considered in detail by Bohr and
Rosenfeld \cite{bohr1}. In this classic paper, a comprehensive
analysis of the measurability of electromagnetic field was given,
and compatibility of the restrictions following from the formal
relations for the field operators with the actual limitations
inherent to any process of measurement was verified, refuting the
earlier arguments by Landau and Peierls \cite{landau} against this
compatibility. In a subsequent paper \cite{bohr2} (see also
\cite{rosenfeld}), the authors argued that the problem of
charge-current measurements can be reduced to that already solved
for the electromagnetic field. Later, a similar investigation was
carried out in quantum theory of gravitation by DeWitt
\cite{dewitt}.

An important point underlying investigation of
Refs.~\cite{bohr1,bohr2} is that the question of measurability of
the electromagnetic field, and hence of charges and currents, can
be considered proceeding from the commutation relations for the
operators describing free electromagnetic field. In particular,
these relations were used to estimate characteristic value of the
electromagnetic field fluctuations, denoted in \cite{bohr1} by
$\mathfrak{S}.$ Defined as the root mean square fluctuation,
$\mathfrak{S}$ is evidently of the order
$O(\sqrt{\hbar}).$\footnote{As was shown in Ref.~\cite{bohr1}, the
value of $\mathfrak{S}$ varies depending on the ratio of the space
and time intervals characterizing the measurement process, but in
any case it is proportional to $\sqrt{\hbar}.$ Below, our main
concern will be the dependence of $\mathfrak{S}$ on the Planck
constant, so we suppress all other factors on which $\mathfrak{S}$
may depend.} It was argued in Ref.~\cite{bohr1} that this estimate
holds true even in the presence of sources, provided that the
charge and current distributions representing these sources allow
classical description (which was also confirmed later by detailed
calculations in Refs.~\cite{thirring,glauber,umezawa,schwinger1}).

As far as one is interested in the electromagnetic field produced
by the test bodies, this result is certainly sufficient to justify
the use of the relation $\mathfrak{S} = O(\sqrt{\hbar}):$ Since
the choice of the experimental setup is at disposal of the
observer, the measuring device may always be assumed
``sufficiently classical.'' However, the state of affairs is
different when the {\it given} electromagnetic field to be
measured is considered. Classical assumption about the field
producing sources is irrelevant in this case, and the above
estimate does not apply. The problem of fundamental importance,
therefore, is to determine fluctuations of fields produced by
nonclassical sources.

On various occasions, this issue has been the subject of a number
of investigations. Electromagnetic field fluctuations springing up
as a response of elementary particles' vacua to external fields or
nontrivial boundary conditions were studied recently in
\cite{hacian,barton,eberlein,ford1,zerbini} where references to
early works can be found. Fluctuations of the gravitational field,
induced by vacuum fluctuations of the matter stress tensor were
considered in Refs.~\cite{zerbini,ford2,ford3,hu1,hu2}.

It should be noted that despite extensive literature in the area,
only vacua contributions of quantized matter fields to the
fluctuations of electromagnetic and gravitational fields have been
studied in detail. At the same time, it is effects produced by
real matter that are most interesting from the point of view of
the structure of elementary contributions to the field
fluctuation.

The purpose of this paper is to investigate fluctuations of the
electromagnetic field produced by a single massive charged
particle, taking full account of quantum properties of the
source-field interaction. It will be shown that the structure of
fluctuations in this case is quite different from that of
vacuum-induced fluctuations, or that found in the case of
classical source. In particular, the root mean square fluctuation
of the field turns out to be of zero order in the Planck constant,
$\mathfrak{S} = O(\hbar^0),$ rather than $O(\sqrt{\hbar}).$ It
seems that this remarkable fact has not been noticed earlier. For
a given electric charge of the source consisting of many
particles, the $\hbar^0$ contribution turns out to be inversely
proportional to the number of constituent particles, so the
classical $O(\sqrt{\hbar})$ estimate is recovered in the
macroscopic limit.

The paper is organized as follows. In Sec.~\ref{prelim} a
preliminary consideration of the problem is given, and the
Schwinger-Keldysh closed time path formalism used throughout the
work is briefly reviewed. General properties of the correlation
function of electromagnetic field fluctuations are considered in
Sec.~\ref{cfproperties}. In investigation of quantum aspects of
particle-field interaction it is convenient to isolate purely
quantum-mechanical effects related to the particle's kinematics
assuming it sufficiently heavy. As discussed in Sec.~\ref{lr},
this leads naturally to the long-range expansion of the
correlation function, making it appropriate to use the terminology
and general ideas of effective field theories
\cite{weinberg1,weinberg2,donoghue1,donoghue2}. Next, it is proved
in Sec.~\ref{gd} that the leading term in the long-range expansion
of the correlation function, which describes fluctuations of the
static potential, is independent of the choice of gauge condition
used to fix the gradient invariance. Finally, the correlation
function is evaluated explicitly in Sec.~\ref{calcul}.
Sec.~\ref{conclud} contains discussion of the results obtained.

\section{Preliminaries}\label{prelim}

Let us consider a single particle with mass $m$ and electric
charge $e.$ In classical theory, the electromagnetic field
produced by such particle at rest is described by the Coulomb
potential
\begin{eqnarray}\label{coulomb}&&
A_0 = \frac{e}{4\pi r}\,, \qquad \bm{A} = 0\,.
\end{eqnarray}
\noindent Our aim is to determine properties of this potential in
quantum domain. For this purpose, it is convenient to separate
this field-theoretic problem from purely quantum mechanical issues
related to the quantum kinematics of the particle. Namely, the
particle's mass will be assumed sufficiently large to neglect the
indeterminacy in the values of particle velocity and position,
following from the Heisenberg principle, so as to still be able to
use the term ``particle at rest.''

In quantum theory, Eq.~(\ref{coulomb}) is reproduced by
calculating the corresponding mean fields, $\langle
\hat{A}_0\rangle,$ $\langle \hat{\bm{A}} \rangle.$ In the
functional integral formalism, these are given by
\begin{eqnarray}\label{fint}
\langle \hat{A}_{\mu}\rangle = \int \EuScript{D}\Phi
A_{\mu}\exp\{i S\}\,,
\end{eqnarray}
\noindent where $\Phi$ collectively denotes the fundamental fields
of the theory, $\Phi = \{A_{\mu},\phi,\phi^*\},$ the components
$\phi,\phi^*$ describing the charged particle which for simplicity
will be assumed scalar, $\EuScript{D}\Phi$ is the invariant
integral measure, and $S = S[\Phi]$ the action functional of the
system. Integration is carried over all field configurations
satisfying
$$A^{\pm}\to 0\,, \quad \phi^{\pm} \to \phi_0^{\pm}\,, \quad
(\phi^*)^{\pm} \to (\phi_0^*)^{\pm}\,, \qquad {\rm for}\quad t \to
\mp\infty\,,$$ where the superscripts $+$ and $-$ denote the
positive- and negative-frequency parts of the fields,
respectively, and $\phi_0$ describes the particle state. Assuming
that the gradient invariance of the theory is fixed by the Lorentz
condition
\begin{eqnarray}\label{gauge}
G\equiv\partial^{\mu}A_{\mu} = 0\,,
\end{eqnarray}
\noindent the action takes the form
\begin{eqnarray}\label{action}
S[\Phi] &=& S_0[\Phi] + S_{\rm gf}[\Phi]\,, \nonumber\\
S_0[\Phi] &=& {\displaystyle\int} d^4 x
\left\{(\partial_{\mu}\phi^* + i e A_{\mu}\phi^*)
(\partial^{\mu}\phi - ie A^{\mu}\phi) - m^2 \phi^*\phi\right\} -
\frac{1}{4}{\displaystyle\int} d^4 x F_{\mu\nu} F^{\mu\nu}\,,
\nonumber\\ S_{\rm gf}[\Phi] &=&
-\frac{1}{2\xi}{\displaystyle\int} d^4 x~G^2\,,\quad F_{\mu\nu} =
\partial_{\mu} A_{\nu} - \partial_{\nu} A_{\mu}\,.
\end{eqnarray}
\noindent where $\xi$ is the Feynman weighting parameter.
Introducing auxiliary classical sources $\EuScript{J} = \{J^{\mu},
j^*,j\}$ for the fields $\Phi = \{A_{\mu},\phi,\phi^*\},$
respectively, the right hand side of Eq.~(\ref{fint}) may be
rewritten in the standard way
\begin{eqnarray}\label{fintj}
\langle \hat{A}_{\mu}\rangle &=& \frac{\delta}{i\delta
J^{\mu}}\exp\left\{S^{\rm{int}}\left[\frac{\delta}{i\delta
\EuScript{J}}\right]\right\}\left.\int
\EuScript{D}\Phi~\exp\left\{i \left(S^{(2)} + \int d^4 x
\left[J^{\mu}A_{\mu} + j\phi^* + j^*\phi\right]
\right)\right\}\right|_{\EuScript{J}=0} \nonumber\\ &=&
\frac{\delta}{i\delta
J^{\mu}}\exp\left\{S^{\rm{int}}\left[\frac{\delta}{i\delta
\EuScript{J}}\right]\right\} \left.\exp \left\{i\int d^4 x\int d^4
y \left[\textstyle{\frac{1}{2}}J^{\mu}(x)D_{\mu\nu}(x,y)J^{\nu}(y)
\right.\right.\right.\nonumber\\&& \left.\left.\left. +
j^*(x)D(x,y)j(y)\right] + i\int d^4 x~(j\phi_0^* +
j^*\phi_0)\right\}\right|_{\EuScript{J}=0}\,,
\end{eqnarray}
\noindent where $S^{(2)}$ denotes the free field part of the
action, $S^{\rm{int}}[\Phi] = S[\Phi] - S^{(2)}[\Phi],$
$D_{\mu\nu}$ and $D$ are the propagators of the electromagnetic
and scalar field, respectively, defined by
\begin{eqnarray}\label{apropord}
\int d^4 z~ \frac{\delta^2S^{(2)}}{\delta A^{\mu}(x)\delta
A^{\nu}(z)}D^{\nu\alpha}(z,y)&=& -
\delta_{\mu}^{\alpha}\delta^{(4)}(x-y)\,,\\
\int d^4 z~ \frac{\delta^2S^{(2)}}{\delta \phi^*(x)\delta
\phi(z)}D(z,y) &=& - \delta^{(4)}(x-y)\,.
\end{eqnarray}
\noindent It is not difficult to verify that the result of
evaluation of the right hand side of Eq.~(\ref{fintj}) in the tree
approximation is given exactly by Eq.~(\ref{coulomb}). The field
fluctuation, however, cannot be determined just as directly. The
point is that the formal expressions like $\langle \hat{A}_0^2(x)
\rangle$ are not well defined because of the singular behavior of
the product $\hat{A}_0(x)\hat{A}_0(y)$ as $x \to y.$ This is a
well known problem, encountered already in the theory of free
fields. As was emphasized in Ref.~\cite{bohr1}, in any field
measurement in a given spacetime point, one deals actually with
the field averaged over a small but finite spacetime domain
surrounding this point, so the physically sensible expression for
the field operator is the following
\begin{eqnarray}\label{stav}
\hat{\EuScript{A}}_{\mu} = \frac{1}{VT}\int_T dt\int_V d^3
\bm{x}~\hat{A}_{\mu}(\bm{x},t)\,.
\end{eqnarray}
\noindent Respectively, the product of two fields in a given point
is understood as the limit of
\begin{eqnarray}\label{stavprod}
\hat{\EuScript{B}}_{\mu\nu} = \frac{1}{(VT)^2}\int_T dt\int_T
dt'\int_V d^3\bm{x} \int_V d^3 \bm{x}'~\hat{A}_{\mu}(\bm{x},t)
\hat{A}_{\nu}(\bm{x}',t')
\end{eqnarray} \noindent when the size of the domain tends to zero.
Finally, the correlation function of the electromagnetic
4-potential in this domain is
\begin{eqnarray}\label{a0av}
\EuScript{C}_{\mu\nu} = \langle \hat{\EuScript{B}}_{\mu\nu}\rangle
- \langle\hat{\EuScript{A}}_{\mu}\rangle
\langle\hat{\EuScript{A}}_{\nu}\rangle\,.
\end{eqnarray}
\noindent

The necessity of spacetime averaging of field operators,
Eq.~(\ref{stav}), dictated by the physical measurement conditions,
entails the following important complication in calculating their
correlation functions. The standard Feynman rules for constructing
matrix elements of a product of field operators, such as that in
the right hand side of Eq.~(\ref{stavprod}), give the in-out
matrix element of the time ordered product of the operators, i.e.,
the Green function, rather than its in-in expectation value. This
circumstance is not essential as long as one is interested in
evaluating (under stationary external conditions) the mean fields.
However, it does make a difference in calculating the correlation
function whether or not the field operators are smeared over a
finite spacetime region. As is well known, in order to find the
expectation value of a product of operators taken in different
spacetime points, the usual Feynman rules for constructing the
matrix elements must be modified. According to the so-called
closed time path formalism \cite{keldysh,schwinger2} (for modern
reviews of the method, see Refs.~\cite{jordan,paz}), the in-in
matrix element of the product $\hat{A}^{\mu}(x)\hat{A}^{\nu}(x')$
can be written as
\begin{eqnarray}\label{fintctp1}
C^{\mu\nu}(x,x') &\equiv&\langle {\rm in}
|\hat{A}^{\mu}(x)\hat{A}^{\nu}(x')|{\rm in}\rangle \nonumber\\ &=&
\int \EuScript{D}\Phi_{-}\int
\EuScript{D}\Phi_{+}~A^{\mu}_-(x)A^{\nu}_+(x')\exp\{i S[\Phi_{+}]
- i S[\Phi_{-}]\}\,,
\end{eqnarray}
\noindent where the subscript $+$ ($-$) shows that the time
argument of the integration variable runs from $-\infty$ to
$+\infty$ (from $+\infty$ to $-\infty$). Integration is over all
fields satisfying
\begin{eqnarray}\label{bcond}
A^{+}_{\pm} \to 0\,, \quad \phi^{+}_{\pm} \to \phi_0^+ \,, \quad
(\phi^*)^{+}_{\pm} \to (\phi^*_0)^+  \qquad {\rm for} \quad t
&\to& -\infty\,, \nonumber\\ \Phi_{+} = \Phi_{-} \qquad {\rm for}
\quad t &\to& + \infty\,.
\end{eqnarray} \noindent It is seen that $\langle {\rm
in} |\hat{A}^{\mu}(x)\hat{A}^{\nu}(x')|{\rm in}\rangle$ is given
by the ordinary functional integral but with the number of fields
doubled, and unusual boundary conditions specified above.
Accordingly, diagrammatics generated upon expanding this integral
in powers of the coupling constant $e$ consists of the following
elements. There are four types of pairings for each field
$A_{\mu}$ or $\phi,$ corresponding to the four different ways of
placing two field operators on the two branches of the time path.
They are conveniently combined into $2 \times 2$
matrices\footnote{Below, Gothic letters are used to distinguish
quantities representing columns, matrices etc. with respect to
indices $+,-.$}
$$\mathfrak{D}^{\mu\nu}(x,y) = \left(
\begin{array}{cc}
D^{\mu\nu}_{++}(x,y)&D^{\mu\nu}_{+-}(x,y)\\
D^{\mu\nu}_{-+}(x,y)&D^{\mu\nu}_{--}(x,y)
\end{array}\right) =
\left(\begin{array}{cc}i\langle
T\hat{A}^{\mu}(x)\hat{A}^{\nu}(y)\rangle_0 & i\langle
\hat{A}^{\nu}(y)\hat{A}^{\mu}(x)\rangle_0\\
i\langle \hat{A}^{\mu}(x)\hat{A}^{\nu}(y)\rangle_0 & i\langle
\tilde{T}\hat{A}^{\mu}(x)\hat{A}^{\nu}(y)\rangle_0\end{array}\right)\,,
$$
$$\mathfrak{D}(x,y) = \left(
\begin{array}{cc}
D_{++}(x,y)&D_{+-}(x,y)\\
D_{-+}(x,y)&D_{--}(x,y)
\end{array}\right) =
\left(\begin{array}{cc}i\langle
T\hat{\phi}(x)\hat{\phi}^\dag(y)\rangle_0 & i\langle
\hat{\phi}(y)\hat{\phi}^\dag(x)\rangle_0\\i\langle
\hat{\phi}(x)\hat{\phi}^\dag(y)\rangle_0 & i\langle
\tilde{T}\hat{\phi}(x)\hat{\phi}^\dag(y)\rangle_0\end{array}\right)\,,
$$
where the operation of time ordering $T$ ($\tilde{T}$) arranges
the factors so that the time arguments decrease (increase) from
left to right, and $\langle\cdot\rangle_0$ denotes vacuum
averaging. The ``propagators'' $\mathfrak{D}_{\mu\nu},$
$\mathfrak{D}$ satisfy the following matrix equations
\begin{eqnarray}\label{aprop}
\int d^4 z~\mathfrak{G}_{\mu\nu}(x,z)\mathfrak{D}^{\nu\alpha}(z,y)
&=& - \mathfrak{e}\delta_{\mu}^{\alpha}\delta^{(4)}(x-y)\,,
\quad\mathfrak{G}_{\mu\nu}(x,y) =
\mathfrak{i}~\frac{\delta^2S^{(2)}}{\delta
A^{\mu}(x)\delta A^{\nu}(y)}\,, \\
\int d^4 z~\mathfrak{G}(x,z)\mathfrak{D}(z,y) &=& -
\mathfrak{e}\delta^{(4)}(x-y)\,, \quad\mathfrak{G}(x,y) =
\mathfrak{i}~\frac{\delta^2S^{(2)}}{\delta \phi^*(x)\delta
\phi(y)}\,,\label{aprop1}
\end{eqnarray}
\noindent where $\mathfrak{e},$ $\mathfrak{i}$ are $2\times2$
matrices with respect to indices $+,-:$
$$\mathfrak{e} = \left(\begin{array}{cc}
1&0\\0&1\end{array}\right)\,, \quad \mathfrak{i} =
\left(\begin{array}{cc} 1&0\\0&-1\end{array}\right)\,.$$ As in the
ordinary Feynman diagrammatics of the S-matrix theory, the
propagators are contracted with the vertex factors generated by
the interaction part of the action, $S^{\rm int}[\Phi],$ with
subsequent summation over $(+,-)$ in the vertices, each ``$-$''
vertex coming with an extra factor $(-1).$ This can be represented
as the matrix multiplication of
$\mathfrak{D}_{\mu\nu},\mathfrak{D}$ with suitable matrix
vertices. For instance, the $A\partial\phi\phi^*$ part of the
action generates the matrix vertex $\mathfrak{V}$ which in
components has the form
$$V_{ijk}^{\mu}(x,y,z) = s_{ijk}\left.\frac{\delta^3 S}{\delta A_{\mu}(x)
\delta\phi(y)\delta\phi^*(z)}\right|_{\Phi = 0}\,,$$ where the
indices $i,j,k$ take the values $+,-,$ and $s_{ijk}$ is defined by
$s_{+++} = - s_{---} = 1$ and zero otherwise. External $\phi$
($\phi^*$) line is represented in this notation by a column (row)
$$\mathfrak{r} = \left(\begin{array}{c} \phi_0 \\ \phi_0
\end{array}\right), \quad \mathfrak{r}^\dag = (\phi^*_0,\phi^*_0),$$ satisfying
\begin{eqnarray}\label{free}
\int d^4 z~\mathfrak{G}(x,z)\mathfrak{r}(z) = \left(\begin{array}{c} 0 \\
0 \end{array}\right), \quad \int d^4 x ~\mathfrak{r}^\dag(x)
\mathfrak{G}(x,z) = (0,0)\,.
\end{eqnarray}
\noindent

Figure \ref{fig1} depicts the tree diagrams contributing to the
right hand side of Eq.~(\ref{fintctp1}). The disconnected part
shown in Fig.~\ref{fig1}(a) cancels in the expression for the
correlation function [see Eq.~(\ref{a0av})], which is thus
represented by the diagrams (b)--(h).

Explicit expressions for various pairings of the photon and scalar
fields
\begin{eqnarray}\label{explprop}
\mathfrak{D}_{\mu\nu} &=& \eta_{\mu\nu}\mathfrak{D}^{0}\,, \qquad
\mathfrak{D}^{0} \equiv \mathfrak{D}|_{m=0}\,, \nonumber\\
D_{++}(x,y) &=& \int\frac{d^4 k}{(2\pi)^4}\frac{e^{-ik(x-y)}}{m^2
- k^2 - i0}\,, \quad D_{--}(x,y) =
\int\frac{d^4 k}{(2\pi)^4}\frac{e^{-ik(x-y)}}{k^2 - m^2 - i0}\,,\nonumber\\
D_{-+}(x,y) &=& i\int\frac{d^4 k}{(2\pi)^3}\theta(k^0)\delta(k^2 -
m^2)e^{-ik(x-y)}\,, \quad D_{+-}(x,y) =  D_{-+}(y,x)\,.
\end{eqnarray}
\noindent The photon propagator is written here in the Feynman
gauge $\xi = 1$ which is most convenient in actual calculations.
The question of how the choice of gauge condition affects the
correlation function is considered in Sec.~\ref{gd}.

\section{Long-range properties of correlation function}\label{cfproperties}

Before we proceed to calculation of the correlation function, we
shall examine its general properties in more detail. Namely, the
structure of the long-range expansion of the correlation function,
and the question of its gauge dependence will be considered.

\subsection{Correlation function in the long-range limit}\label{lr}

The mean electromagnetic field produced by a massive charged
particle is a function of five dimensional parameters - the
fundamental constants $\hbar,c,$ the charge $e$ and mass $m$ of
the particle, and the distance between the particle and the point
of observation, $r.$ Of these only two independent dimensionless
combinations can be constructed -- the constant $e^2/\hbar c$
playing the role of the expansion parameter of perturbation
theory, and the ratio $l_c/r \equiv \varkappa\,,$ where $l_c =
\hbar/mc$ is the Compton length of the particle. As we have
mentioned above, the particle is assumed sufficiently heavy so as
to neglect effects related to the particle kinematics. This means
that the quantum fluctuations are investigated in the limit
$\varkappa \to 0.$ For fixed particle's mass this implies large
values of $r.$ In other words, the relevant information about
field correlations is contained in the long-range behavior of the
quantity $\EuScript{C}_{\mu\nu}.$

To extract this information we note, first of all, that in the
long-range limit, the value of $\EuScript{C}_{\mu\nu}$ is
independent of the choice of spacetime domain used in the
definition of physical electromagnetic field operators,
Eq.~(\ref{stav}). Indeed, in any case the size of this domain must
be small in comparison with the characteristic length at which the
mean field changes significantly. In the case considered, this
means that $V \ll r^3.$ To the leading order of the long-range
expansion, therefore, the quantity $\langle\hat{A}_{\mu}(x)
\hat{A}_{\nu}(x')\rangle$ appearing in the right hand side of
Eq.~(\ref{a0av}) can be considered constant within the domain.
However, one cannot set $x=x'$ in this expression directly. It is
not difficult to see that the formal expression
$\langle\hat{A}_{\mu}(x) \hat{A}_{\nu}(x)\rangle$ does not exist.
Consider, for instance, the diagram \ref{fig1}(b). It is
proportional to the integral $$I_{\mu\nu}(x-x',p) = \int d^4 k
\theta(k^0)\delta(k^2)e^{ik(x-x')}\frac{(2q_{\mu} + k_{\mu} +
p_{\mu})(2q_{\nu} + k_{\nu}) }{[(k+q)^2 - m^2](k-p)^2}\ ,$$ where
$q_{\mu}$ is the 4-momentum of the scalar particle, and $p_{\mu}$
the momentum transfer. For small but nonzero $(x-x')$ this
integral is effectively cut-off at large $k$'s by the oscillating
exponent, but for $x=x'$ it is divergent. This divergence arises
from integration over large values of virtual photon momenta, and
therefore has nothing to do with the long-range behavior of the
correlation function, because this behavior is determined by the
low-energy properties of the theory. Evidently, the singularity of
$I_{\mu\nu}(x-x',p)$ for $x \to x'$ is not worse than $\ln(x^{\mu}
- x^{\prime\mu})^{2}/(x^{\mu} - x^{\prime\mu})^{2}.$ Therefore,
$\langle {\rm in} |\hat{A}_{\mu}(x)\hat{A}_{\nu}(x')|{\rm
in}\rangle$ is integrable, and $\EuScript{C}_{\mu\nu}$ given by
Eqs.~(\ref{stavprod}), (\ref{a0av}) is well defined.

Our aim below will be to show that this singularity can be
consistently isolated and removed from the expression for
$I_{\mu\nu}(x-x',p)$ and similar integrals for the rest of
diagrams in Fig.~\ref{fig1}, without changing  the long-range
properties of the correlation function. After this removal, it is
safe to set $x=x'$ in the finite remainder, and to consider
$\EuScript{C}_{\mu\nu}$ as a function of the single variable --
the distance $r.$ An essential point of this procedure is that the
singularity turns out to be {\it local}, and hence does not
interfere with terms describing the long-range behavior, which
guaranties unambiguity of the whole procedure.

It should be emphasized that in contrast to what takes place in
the scattering theory, the ultraviolet divergences appearing in
the course of calculation of the in-in matrix elements in the
coincidence limit are generally non-polynomial with respect to the
momentum transfer. The reason for this is the different analytic
structure of various elements in the matrix propagators
$\mathfrak{D}_{\mu\nu},$ $\mathfrak{D},$ which spoils the simple
ultraviolet properties exhibited by the ordinary Feynman
amplitudes.\footnote{As is well known, the proof of locality of
the S-matrix divergences relies substantially on the causality of
the pole structure of Feynman propagators. This property allows
Wick rotation of the energy contours, thus revealing the
essentially Euclidean nature of the ultraviolet divergences.} Take
the above integral as an example. Because of the delta function in
the integrand, differentiation of $I_{\mu\nu}(x-x',p)$ with
respect to the momentum transfer does not remove the ultraviolet
divergence of $I_{\mu\nu}(0,p).$ What makes it all the more
interesting is the result obtained in Sec.~\ref{calcul} below,
that the non-polynomial parts of divergent contributions
eventually cancel each other, and the overall divergence turns out
to be completely local.

Next, let us establish general form of the leading term in the
long-range expansion of the correlation function. In momentum
representation, an expression of lowest order in the momentum
transfer with suitable dimension and Lorentz transformation
properties is the following
\begin{eqnarray}\label{zc}
\EuScript{C}^{(0)}_{\mu\nu}(p) \sim
\frac{e^2}{m^2}\frac{q_{\mu}q_{\nu}}{\sqrt{-p^2}}\ .
\end{eqnarray}
\noindent Thus, only 00-component of the correlation function
survives in the long-range limit: $$\EuScript{C}^{(0)}_{00}(p)
\sim \frac{e^2}{\sqrt{-p^2}}\ .$$ Not all diagrams of
Fig.~\ref{fig1} contain contributions of this type. It is not
difficult to identify those which do not. Consider, for instance,
the diagram (h). It is proportional to the integral
$$\int d^4 k
\theta(k^0)\delta(k^2)\frac{e^{ik(x-x')}}{(k-p)^2}\ ,$$ which does
not involve the particle mass at all. Taking into account that
each external scalar line gives rise to the factor
$(2\varepsilon_{\bm{q}})^{-1/2},$ where $\varepsilon_{\bm{q}} =
\sqrt{m^2 + \bm{q}^2}\approx m\,,$ we see that the contribution of
the diagram (h) is proportional to $1/m.$ The same is true of all
other diagrams without internal scalar lines. As to diagrams
involving such lines, it will be shown in Sec.~\ref{calcul} by
direct calculation that they do contain contributions of the type
Eq.~(\ref{zc}). But prior to this the question of their dependence
on the gauge will be considered.

\subsection{Gauge independence of the leading
contribution}\label{gd}

The value of 0-component of the momentum transfer is fixed by the
mass shell condition for the scalar particle: $$p^0 = \sqrt{m^2 +
(\bm{p} + \bm{q})^2} - \sqrt{m^2 + \bm{q}^2} =
O\left(\frac{|\bm{p}|}{m}\right)\,.$$ It follows from this
estimate that $p^0$ is to be set zero in the long-range limit.
This implies that $\EuScript{C}^{(0)}_{\mu\nu}$ represents
fluctuations in a quantity of direct physical meaning -- the
static potential energy of interacting particles. As such it is
expected to be independent of the gauge condition used to fix the
gradient invariance. More precisely, the issue of gauge dependence
in the present case consists in the following. Although the gauge
condition (\ref{gauge}) fixes the form of the classical solution
(\ref{coulomb}) unambiguously, the inverse is not true: There are
infinitely many ways of constructing the gauge fixed action, which
lead to the same expression for the static potential. We have to
verify that this freedom in the choice of the gauge condition does
not affect the value of $\EuScript{C}^{(0)}_{\mu\nu}.$

The gauge independence of the leading contribution will be
demonstrated below in the most important particular case when
changes of the gauge-fixed action are induced by variations of the
Feynman weighting parameter $\xi.$ It is not difficult to see that
these variations do not alter the value of classical potential.
Indeed, the latter satisfies
\begin{eqnarray}\label{class}
\frac{\delta S_{\rm em}}{\delta A_{\mu}} = - J^{\mu}\,,
\end{eqnarray}
\noindent where $S_{\rm em}$ is the gauge-fixed action for
electromagnetic field, $$ S_{\rm em} = -
\frac{1}{4}{\displaystyle\int} d^4 x \left(F_{\mu\nu} F^{\mu\nu} +
\frac{2G^2}{\xi}\right)\,,$$ and $J^{\mu}$ the electromagnetic
current, $$J^{\mu} = ie\langle {\rm in} |
\hat{\phi}^\dag\partial^{\mu}\hat{\phi} -
\hat{\phi}\partial^{\mu}\hat{\phi}^\dag |{\rm in}\rangle\,.$$
Acting on Eq.~(\ref{class}) with $\partial_{\mu},$ using the
current conservation $\partial_{\mu}J^{\mu} = 0$ and the identity
$\partial_{\mu}\partial_{\nu}F^{\mu\nu} = 0,$ we obtain
$$\frac{1}{\xi}\Box G = 0\,,$$ which implies that $G=0$ on the
classical solution, and that $\xi$ effectively falls off from the
classical equation (\ref{class}). Let us now prove that
$\EuScript{C}^{(0)}_{\mu\nu}$ remains unchanged under variations
of $\xi$ too. We shall proceed along the lines of
Ref.~\cite{kazakov1} where $\xi$-independence of $\hbar^0$ part of
the one-loop correction to Newtonian potential was proved within
the S-matrix approach.

As we saw in Sec.~\ref{lr}, the only diagrams contributing in the
long-range limit are those containing internal scalar lines. We
will show presently that the $\xi$-dependent part of these
diagrams can be reduced to the form without such lines. First, it
follows form Eq.~(\ref{aprop}) that the $\xi$-derivative of the
electromagnetic propagator satisfies
\begin{eqnarray}\label{xideriv}
\frac{\partial \mathfrak{D}_{\mu\nu}(x,y)}{\partial\xi} = \iint
d^4 z d^4
z'~\mathfrak{D}_{\mu\alpha}(x,z)\frac{\partial\mathfrak{G}^{\alpha\beta}(z,z')}
{\partial\xi}\mathfrak{D}_{\beta\nu}(z',y)\,.
\end{eqnarray}
\noindent On the other hand, acting on Eq.~(\ref{aprop}) by
$\partial^{\mu}$ one finds
\begin{eqnarray}\label{stderiv}
\frac{1}{\xi}I\Box_{x}\partial_{x}^{\nu}\mathfrak{D}_{\nu\alpha}(x,y)
= - E\partial^{x}_{\alpha}\delta^{(4)}(x-y)\,.
\end{eqnarray}
\noindent This equation can be resolved with respect to
$\partial^{\nu}\mathfrak{D}_{\nu\alpha}.$ The reciprocal of the
operator $I\Box,$ appropriate to the boundary conditions
(\ref{bcond}) is $(-\mathfrak{D}^0),$ hence
\begin{eqnarray}\label{solv}
\partial_{x}^{\nu}\mathfrak{D}_{\nu\alpha}(x,y) =
\xi \partial^{x}_{\alpha}\mathfrak{D}^0(x,y)\,.
\end{eqnarray}
\noindent Taking into account that
$$\frac{\partial\mathfrak{G}^{\alpha\beta}(z,z')}{\partial\xi} = -
\frac{I}{\xi^2}\partial_{z}^{\alpha}\partial_{z}^{\beta}\delta^{(4)}(z-z')\,,$$
substituting this into Eq.~(\ref{xideriv}), and using
Eq.~(\ref{solv}), we get
\begin{eqnarray}
\frac{\partial \mathfrak{D}_{\mu\nu}(x,y)}{\partial\xi} = -
\frac{1}{\xi}\int d^4
z~\mathfrak{D}_{\mu\alpha}(x,z)I\partial_z^{\alpha}
\partial^z_{\nu}\mathfrak{D}^0(z,y)\,, \nonumber
\end{eqnarray}
\noindent or
\begin{eqnarray}\label{xideriv1}
\frac{\partial \mathfrak{D}_{\mu\nu}(x,y)}{\partial\xi} =
\partial^x_{\mu}\partial^y_{\nu}\mathfrak{a}(x,y)\,, \qquad
\mathfrak{a}(x,y)\equiv \int d^4
z~\mathfrak{D}^0(x,z)I\mathfrak{D}^0(z,y)\,.
\end{eqnarray}
\noindent Thus, assuming the matrix indices and spacetime
coordinates referring to the point of observation fixed, we see
that the $\xi$-dependent part of the matrix photon propagator is
contracted with the matrix vertex $\mathfrak{V}^{\mu}$ through a
gradient term
$$\partial_{\mu}\mathfrak{b}(x) =
\left(\begin{array}{c} \partial_{\mu}b_+(x) \\
\partial_{\mu}b_-(x)
\end{array}\right)\,, \qquad \mathfrak{b}(x) =
\partial^y_{\nu}\mathfrak{a}(x,y)\,.$$ Integrating by parts
the spacetime derivative may be rendered to act on the vertex.

Next, we use gauge invariance of the action $S_0[\Phi]$ to
transform $\mathfrak{b}\partial_{\mu}\mathfrak{V}^{\mu}.$ This
invariance is expressed by the identity
\begin{eqnarray}\label{ident}
\frac{\delta S_0}{\delta\phi(x)}ie\phi(x) - \frac{\delta
S_0}{\delta\phi^*(x)}ie\phi^*(x) -
\partial^x_{\mu}\frac{\delta S_0}{\delta A_{\mu}(x)} = 0\,.
\end{eqnarray}
\noindent Differentiating Eq.~(\ref{ident}) twice with respect to
$\phi,\phi^*,$ and setting $\Phi = 0$ gives
\begin{eqnarray}\label{ident1}
\hspace{-1cm}\left.\partial^x_{\mu}\frac{\delta^3 S_0}{\delta
A_{\mu}(x)\delta\phi(y)\delta\phi^*(z)}\right|_{\Phi = 0} &=&
\frac{\delta^2
S^{(2)}_0}{\delta\phi^*(z)\delta\phi(x)}ie\delta^{(4)}(x-y) -
\frac{\delta^2
S^{(2)}_0}{\delta\phi(y)\delta\phi^*(x)}ie\delta^{(4)}(x-z)\,.
\end{eqnarray}
\noindent The matrix vertex $\mathfrak{V}^{\mu}$ is obtained by
multiplying $\delta^3 S_0 /\delta\phi\delta\phi^*\delta A_{\mu}$
by the matrix $s_{ijk}.$ It follows from Eq.~(\ref{ident1}) that
the combination $\mathfrak{b}\partial_{\mu}\mathfrak{V}^{\mu}$ may
be written as
\begin{eqnarray}\label{ident2}
b_i\partial^x_{\mu}V_{ijk}^{\mu}(x,y,z) &=& ie(b_+s_{+jk} +
b_-s_{-jk})\left\{\frac{\delta^2
S^{(2)}_0}{\delta\phi^*(z)\delta\phi(x)} \delta^{(4)}(x-y) -
\frac{\delta^2
S^{(2)}_0}{\delta\phi(y)\delta\phi^*(x)}\delta^{(4)}(x-z)\right\},
\nonumber
\end{eqnarray}
\noindent or
\begin{eqnarray}\label{ident3}
\mathfrak{b}\partial^x_{\mu}\mathfrak{V}^{\mu}(x,y,z) =
ie\left\{\mathfrak{F}(z,x)\delta^{(4)}(x-y) - \mathfrak{F}(x,y)
\delta^{(4)}(x-z)\right\}\,,
\end{eqnarray}
\noindent where $$\mathfrak{F}(x,y) = \mathfrak{f}~\frac{\delta^2
S^{(2)}_0}{\delta\phi^*(x)\delta\phi(y)}\,, \qquad \mathfrak{f} =
\left(\begin{array}{cc} b_+&0\\0&-b_-\end{array}\right)\,.$$

Finally, contracting
$\mathfrak{b}\partial_{\mu}\mathfrak{V}^{\mu}(x,y,z)$ with the
matrix $\mathfrak{D}(y,y')$ and the vector
$\tilde{\mathfrak{r}}(z) = \left(\phi_0(z),\phi_0(z)\right)\,,$
and using Eqs.~(\ref{aprop1}), (\ref{free}) gives
\begin{eqnarray}
\iint d^4 y d^4
z~\tilde{\mathfrak{r}}(z)\mathfrak{b}\partial^x_{\mu}
\mathfrak{V}^{\mu}(x,y,z)\mathfrak{D}(y,y') &=& ie\int d^4
z~\tilde{\mathfrak{r}}(z)\mathfrak{F}(z,x)~\mathfrak{D}(x,y') \nonumber\\
- ie\tilde{\mathfrak{r}}(x)\int d^4 y
\mathfrak{F}(x,y)\mathfrak{D}(y,y') &=&
ie\phi_0\tilde{\mathfrak{b}}\delta^{(4)}(x-y')\,. \nonumber
\end{eqnarray}
\noindent We see that upon extracting the $\xi$-dependent part of
the diagrams (b)--(f) in Fig.~\ref{fig1}, the scalar propagator is
cancelled by the vertex factor. Using the reasoning of
Sec.~\ref{lr}, the resulting diagrams give rise to terms $\sim
1/m\,.$ Thus, $\xi$-independence of the correlation function in
the long-range limit is proved.

\section{Evaluation of the leading contribution}
\label{calcul}

\begin{figure}
\includegraphics{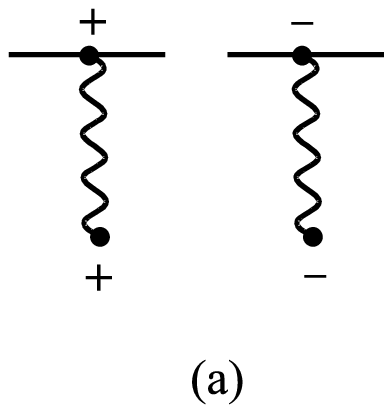}
\includegraphics{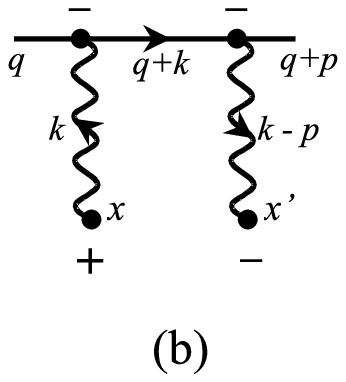}
\includegraphics{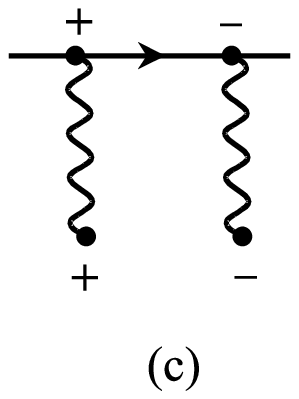}
\includegraphics{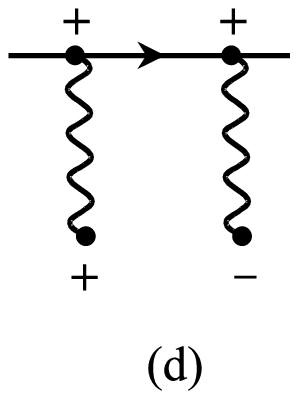}
\includegraphics{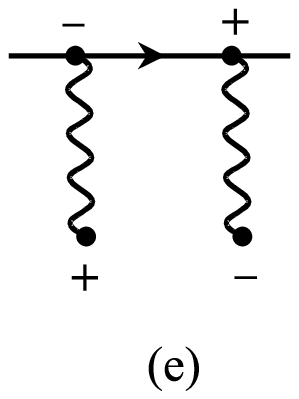}
\includegraphics{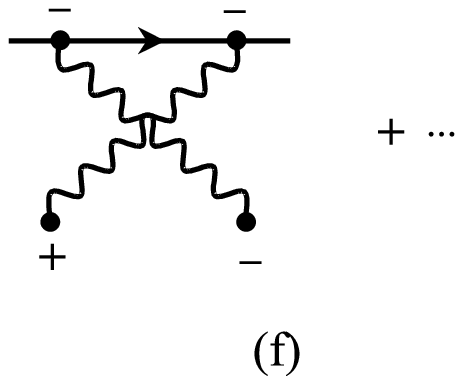}\\
\includegraphics{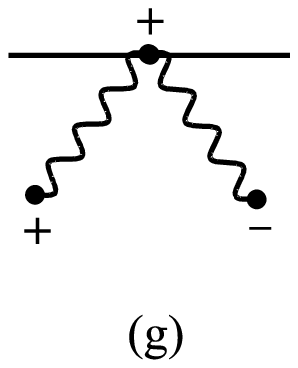}
\includegraphics{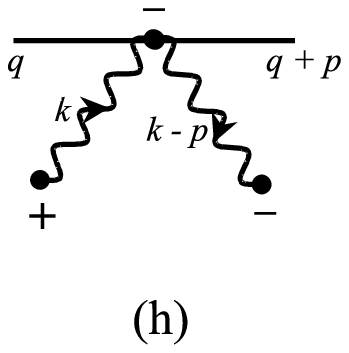} \caption{Tree
contribution to the right hand side of Eq.~(\ref{fintctp1}). Wavy
lines represent photon propagators, solid lines scalar particle.
Part (f) of the figure represents the ``transposition'' of
diagrams (b)--(e) (see Sec.~\ref{calcul}).} \label{fig1}
\end{figure}

Let us proceed to calculation of the leading contribution to the
correlation function. It is contained in the sum of diagrams
(b)--(f) in Fig.~\ref{fig1}, which has the symbolic form
$$C_{\mu\nu} = I_{\mu\nu} + I^{\rm tr}_{\mu\nu}\,, \qquad I_{\mu\nu} =
\frac{1}{i}\left\{\mathfrak{D}_{\mu\alpha}\left[\mathfrak{r}^\dag
\mathfrak{V}^{\alpha}\mathfrak{D}\mathfrak{V}^{\beta}\mathfrak{r}\right]
\mathfrak{D}_{\beta\nu}\right\}_{+-}\,,$$ where the superscript
``tr'' means transposition of the indices and spacetime arguments
referring to the points of observation: $\mu\leftrightarrow\nu,$
$+\leftrightarrow -,$ $x\leftrightarrow x'$ [the transposed
contribution is represented by the diagrams collected in part (f)
of Fig.~\ref{fig1}]. As it follows from the considerations of
Sec.~\ref{lr}, $\EuScript{C}^{(0)}_{\mu\nu}$ can be expressed
through $C_{\mu\nu}$ as
\begin{eqnarray}\label{c0i0}
\EuScript{C}^{(0)}_{\mu\nu} =
\lim\limits_{\genfrac{}{}{0pt}{}{m\to\infty}{V,T\to 0
}}\left\{\frac{1}{(VT)^2}\iint\limits_{(V,T)} d^4 x d^4 x'~
C_{\mu\nu}(x,x')\right\}\,.
\end{eqnarray}
\noindent Written longhand, $I_{\mu\nu}$ reads
\begin{eqnarray}\label{diagen}
I_{\mu\nu}(x,x') = ie^2\iint d^4 z d^4 z'\Biggl\{&+&
D^0_{++}(x,z)\left[\phi^*(z)
\stackrel{\leftrightarrow}{\partial_{\mu}}
D_{++}(z,z')\stackrel{\leftrightarrow}{\partial_{\nu}^{\,\prime}}
\phi(z')\right]D^0_{+-}(z',x') \nonumber\\ &-&
D^0_{++}(x,z)\left[\phi^*(z)
\stackrel{\leftrightarrow}{\partial_{\mu}}
D_{+-}(z,z')\stackrel{\leftrightarrow}{\partial_{\nu}^{\,\prime}}
\phi(z')\right]D^0_{--}(z',x')\nonumber\\ &-&
D^0_{+-}(x,z)\left[\phi^*(z)
\stackrel{\leftrightarrow}{\partial_{\mu}}
D_{-+}(z,z')\stackrel{\leftrightarrow}{\partial_{\nu}^{\,\prime}}
\phi(z')\right]D^0_{+-}(z',x')\nonumber\\ &+&
D^0_{+-}(x,z)\left[\phi^*(z)
\stackrel{\leftrightarrow}{\partial_{\mu}}
D_{--}(z,z')\stackrel{\leftrightarrow}{\partial_{\nu}^{\,\prime}}
\phi(z')\right]D^0_{--}(z',x') \Biggr\}\,,\nonumber\\
\end{eqnarray}
\noindent where
$$\varphi\stackrel{\leftrightarrow}{\partial_{\mu}}\psi =
\varphi\partial_{\mu}\psi - \psi\partial_{\mu}\varphi\,.$$
Contribution of the third term in the right hand side of
Eq.~(\ref{diagen}) is zero identically. Indeed, using
Eq.~(\ref{explprop}), and performing spacetime integrations we see
that the three lines coming, say, into $z$-vertex are all on the
mass shell, which is inconsistent with the momentum conservation
in the vertex. The remaining terms in Eq.~(\ref{diagen}) take the
form
\begin{eqnarray}\label{diagenk1}
I_{\mu\nu}(x,x') &=& e^2\iint \frac{d^3 \bm{q}}{(2\pi)^3}
\frac{d^3 \bm{p}}{(2\pi)^3}\frac{a^*(\bm{q})a(\bm{q} +
\bm{p})}{\sqrt{2\varepsilon_{\bm q}2\varepsilon_{{\bm q} +
\bm{p}}}} e^{-ipx'}\tilde{I}_{\mu\nu}(p,q)\,, \quad p_0 =
\varepsilon_{\bm{q} + \bm{p}} - \varepsilon_{\bm{q}}\,,
\end{eqnarray}
\noindent where
\begin{eqnarray}\label{diagenk2}
\tilde{I}_{\mu\nu}(p,q) = - i\int \frac{d^4 k}{(2\pi)^4}&
e^{-ik(x-x')}(2q_{\mu} + k_{\mu})(2q_{\nu} + k_{\nu} +
p_{\nu})&\nonumber\\
\times\Bigl\{&\;\; D^0_{++}(k)D_{++}(q+k)D^0_{+-}(k-p)& \nonumber\\
&-D^0_{++}(k)D_{+-}(q+k)D^0_{--}(k-p)&\nonumber\\
&+D^0_{+-}(k)D_{--}(q+k)D^0_{--}(k-p)&\Bigr\}\,,
\end{eqnarray}
\noindent and $a(\bm{q})$ is the Fourier transform of the particle
wave function, normalized by $$\int\frac{d^3
\bm{q}}{(2\pi)^3}|a(\bm{q})|^2 = 1\,.$$ The function $a(\bm{q})$
is generally of the form
$$a(\bm{q}) = b(\bm{q}) e^{-i\bm{q}\bm{x}_0},$$ where $\bm{x}_0$
is the mean particle position, and $b(\bm{q})$ is such that
\begin{eqnarray}\label{packet}
\int d^3\bm{q}~b(\bm{q})e^{i\bm{qx}} = 0
\end{eqnarray}
\noindent for $\bm{x}$ outside of some finite region $W$ around
$\bm{x} = 0.$

Upon extraction of the leading contribution these expressions
considerably simplify. First of all, one has $\varepsilon_{\bm{q}
+ \bm{p}}\approx \varepsilon_{\bm{q}} \approx m,$ and hence, $p_0
\approx 0.$ Next, $b(\bm{q} + \bm{p})$ may be substituted by
$b(\bm{q}):$ This implies that we disregard spatial spreading of
the wave packet, neglecting the multipole moments of the charge
distribution. Taking into account the normalization condition,
Eq.~(\ref{diagenk1}) thus becomes
\begin{eqnarray}\label{diagenk3}
I_{\mu\nu}(x,x') &=& \frac{e^2}{2m}\iint\frac{d^3
\bm{q}}{(2\pi)^3}\frac{d^3 \bm{p}}{(2\pi)^3}|b(\bm{q})|^2
e^{i\bm{p}(\bm{x'} - \bm{x}_0 )}\tilde{I}_{\mu\nu}(p,q)\,, \quad
p_0 = 0\,.
\end{eqnarray}
\noindent The leading term in $\tilde{I}_{\mu\nu}(p,q)$ has the
form [Cf. Eq.~(\ref{zc})]
\begin{eqnarray}\label{leading}
\tilde{I}^{(0)}_{\mu\nu}(p,q) \sim
\frac{e^2}{m^2}\frac{q_{\mu}q_{\nu}}{\sqrt{-p^2}}\ .
\end{eqnarray}
\noindent This singular at $p\to 0$ contribution comes from
integration over small $k$ in Eq.~(\ref{diagenk2}). Therefore, to
the leading order, the momenta $k,p$ in the vertex factors can be
neglected in comparison with $q.$ Thus,
\begin{eqnarray}\label{diagenk4}
\tilde{I}_{\mu\nu}(p,q) = - 4q_{\mu}q_{\nu}i\int\frac{d^4
k}{(2\pi)^4} e^{-ik(x-x')}
\Bigl\{&&D^0_{++}(k)D_{++}(q+k)D^0_{+-}(k-p) \nonumber\\
&-&D^0_{++}(k)D_{+-}(q+k)D^0_{--}(k-p)\nonumber\\
&+&D^0_{+-}(k)D_{--}(q+k)D^0_{--}(k-p)\Bigr\}\,.
\end{eqnarray}
\noindent Furthermore, it is convenient to combine various terms
in this expression with the corresponding terms in the transposed
contribution. Noting that the right hand side of
Eq.~(\ref{diagenk4}) is explicitly symmetric in $\mu,\nu,$ and
that the variables $x,x'$ in the exponent can be freely
interchanged because they appear symmetrically in
Eq.~(\ref{c0i0}), we may write
\begin{eqnarray}\label{diagenk5}
\tilde{C}_{\mu\nu}(p,q) &\equiv& \tilde{I}_{\mu\nu}(p,q) +
\tilde{I}^{\rm tr}_{\mu\nu}(p,q) = - 4q_{\mu}q_{\nu}i\int
\frac{d^4 k}{(2\pi)^4} e^{-ik(x-x')}\nonumber\\
\times\Bigl\{&\;&\left[D^0_{++}(k)D_{++}(q+k)D^0_{+-}(k-p)
+ D^0_{--}(k)D_{--}(q+k)D^0_{-+}(k-p)\right]\nonumber\\
&+& \left[D^0_{+-}(k)D_{--}(q+k)D^0_{--}(k-p) +
D^0_{-+}(k)D_{++}(q+k)D^0_{++}(k-p)
\right]\nonumber\\
&-& \left[D^0_{++}(k)D_{+-}(q+k)D^0_{--}(k-p) +
D^0_{--}(k)D_{-+}(q+k)D^0_{++}(k-p)\right]\Bigr\}\,.
\end{eqnarray}
\noindent With the help of the relation
\begin{eqnarray}\label{aux2}
D_{--} + D_{++} = D_{+-} + D_{-+}\,,
\end{eqnarray}
\noindent which is a consequence of the identity
$$T\hat{\phi}(x)\hat{\phi}(y) + \tilde{T}\hat{\phi}(x)\hat{\phi}(y)
- \hat{\phi}(x)\hat{\phi}(y) - \hat{\phi}(y)\hat{\phi}(x) = 0\,,$$
the first term in the integrand can be transformed as
\begin{eqnarray}&&
D^0_{++}(k)D_{++}(q+k)D^0_{+-}(k-p) +
D^0_{--}(k)D_{--}(q+k)D^0_{-+}(k-p) \nonumber\\&& =
D^0_{++}(k)D_{++}(q+k)D^0(k-p) +
D^0(k)D_{--}(q+k)D^0_{-+}(k-p)\,,\nonumber
\end{eqnarray}
\noindent where $$D(k) \equiv D_{+-}(k) + D_{-+}(k) = 2\pi i
\delta(k^2 - m^2)\,, \quad D^0(k) = D(k)|_{m=0}\,.$$ Here we used
the already mentioned fact that $D^0_{+-}(k)D_{-+}(k+q) \equiv 0$
for $q$ on the mass shell. Analogously, the second term becomes
$$D^0(k)D_{--}(q+k)D^0_{--}(k-p) - D^0_{-+}(k)D_{--}(q+k)D^0(k-p)\,.$$
Changing the integration variables $k \to k + p,$ $\bm{q} \to
\bm{q} - \bm{p},$ and then $\bm{p} \to - \bm{p},$ noting that the
leading term is even in the momentum transfer [see
Eq.~(\ref{leading})], and that $\bm{p}(\bm{x}-\bm{x}')$ in the
exponent can be omitted in the coincidence limit ($x\to x'$), the
sum of the two terms takes the form
\begin{eqnarray}
D^0(k)\left[D_{++}(q+k)D^0_{++}(k-p) +
D_{--}(q+k)D^0_{--}(k-p)\right]\,. \nonumber
\end{eqnarray}
\noindent Similar transformations of the rest of the integrand
yield
\begin{eqnarray}\label{diagenk6}&&
D^0_{++}(k)D_{+-}(q+k)D^0_{--}(k-p) +
D^0_{--}(k)D_{-+}(q+k)D^0_{++}(k-p) \nonumber\\&&\to -
\frac{1}{2}\left[D^0_{++}(k)D(q+k)D^0_{++}(k-p) +
D^0_{--}(k)D(q+k)D^0_{--}(k-p)\right]\,.
\end{eqnarray}
\noindent Substituting these expressions into Eq.~(\ref{diagenk5})
and using Eq.~(\ref{explprop}) gives
\begin{eqnarray}\label{diagenk61}
\tilde{C}_{\mu\nu}(p,q) &=& 4q_{\mu}q_{\nu}\int\frac{d^4
k}{(2\pi)^3} e^{-ik(x-x')}\nonumber\\ &\times&{\rm Re}\left\{
\frac{2\delta(k^2)}{[(k-p)^2 + i0][(q+k)^2 - m^2 + i0]} +
\frac{\delta[(q+k)^2 - m^2]}{[k^2 + i0][(k-p)^2 + i0]}\right\}\,.
\end{eqnarray}
\noindent As was discussed in Sec.~\ref{lr}, the exponent in the
integrand in Eq.~(\ref{diagenk61}) plays the role of an
ultraviolet cutoff, ensuring convergence of the integral at large
$k$'s. On the other hand, the leading contribution (\ref{leading})
is determined by integrating over $k \sim p$ where it is safe to
take the limit $x \to x'.$ Since $(x-x')$ is eventually set equal
to zero, one can further simplify the $k$ integral by using the
dimensional regulator instead of the oscillating exponent. Namely,
introducing the dimensional regularization of the $k$ integral,
one may set $x=x'$ afterwards to obtain
\begin{eqnarray}\label{diagenk7}
\tilde{C}_{\mu\nu}(p,q) &=& 4q_{\mu}q_{\nu}\mu^{\epsilon}~{\rm
Re}\int \frac{d^{4-\epsilon} k}{(2\pi)^3} \left\{
\frac{2\delta(k^2)}{[(k-p)^2 + i0][(q+k)^2 - m^2 + i0]}
\right.\nonumber\\&+& \left. \frac{\delta[(q+k)^2 - m^2]}{[k^2 +
i0][(k-p)^2 + i0]}\right\}\,,
\end{eqnarray}
\noindent where $\mu$ is an arbitrary mass parameter, and
$\epsilon = 4 - d,$ $d$ being the dimensionality of spacetime.

Next, going over to the $\alpha$-representation, the first term in
the integrand may be parameterized as
\begin{eqnarray}&&
\frac{\delta(k^2)}{[(k-p)^2 + i0][(q+k)^2 - m^2 + i0]} =
\frac{\delta(k^2)}{(p^2 - 2kp + i0)(2kq + i0)} \nonumber\\&& =
\frac{1}{2\pi i^2}\iiint_{0}^{\infty}dx dy dz \left(e^{ixk^2} +
e^{-ixk^2}\right)e^{iy(p^2 - 2kp + i0)}e^{iz(2kq + i0)}
\end{eqnarray}
\noindent Substituting this into Eq.~(\ref{diagenk7}), and using
the formulas $$\int d^{4-\epsilon}k~e^{i(a k^2 + 2bk)} = {\rm
sign}(a)\frac{1}{i}\left(\frac{\pi}{|a|}\right)^{2 -
\epsilon/2}\exp\left(\frac{b^2}{ia}\right),$$
$$\int_{0}^{\infty}dx~x^{- \epsilon} e^{ixa} =
i~\Gamma(1-\epsilon)\exp\left({\rm sign
}(a)\frac{\pi\epsilon}{2i}\right)\frac{|a|^\epsilon}{a}\,,$$ one
finds
\begin{eqnarray}\label{int1}
K(p) &\equiv& \mu^{\epsilon}\int \frac{d^{4-\epsilon}
k}{(2\pi)^3}\frac{\delta(k^2)}{[(k-p)^2 + i0][(q+k)^2 - m^2]}
\nonumber\\ &=&
\frac{i\mu^\epsilon}{(2\pi)^4}\iiint_{0}^{\infty}dx dy dz
\left(\frac{\pi}{x}\right)^{2 -
\epsilon/2}e^{iyp^2}\left\{\exp\left[- \frac{i}{x}(y p - z
q)^2\right] - \exp\left[\frac{i}{x}(y p - z q)^2\right]\right\}
\nonumber\\ &=& \frac{\mu^\epsilon\pi^{-\epsilon/2}}{8\pi^2}
\cos\left(\frac{\pi\epsilon}{4}\right) \Gamma\left(1 -
\frac{\epsilon}{2}\right)\iint_{0}^{\infty} dy
dz~e^{iyp^2}\frac{\left|(y p - z q)^2\right|^{\epsilon/2}}{(y p -
z q)^2}\,.\nonumber
\end{eqnarray}\noindent Changing the integration variable $z\to
yz,$ and taking into account that $q^2 = m^2,$ $qp = -p^2/2$
yields
\begin{eqnarray}\label{int2}
K(p) &=& \frac{(\mu m)^\epsilon\pi^{-\epsilon/2}}{8\pi^2
m^2}\cos\left(\frac{\pi\epsilon}{4}\right) \Gamma\left(1 -
\frac{\epsilon}{2}\right)\int_{0}^{\infty}dy~e^{i y p^2}
y^{\epsilon - 1}\int_{0}^{\infty}dz \frac{\left|z^2 -
(1+z)\alpha\right|^{\epsilon/2}}{z^2 - (1+z)\alpha} \nonumber\\
&=& \frac{(\pi e^{i\pi})^{-\epsilon/2}}{8\pi^2 m^2}\left(\frac{\mu
m}{- p^2}\right)^{\epsilon}\cos\left(\frac{\pi\epsilon}{4}\right)
\Gamma\left(1 - \frac{\epsilon}{2}\right)
\Gamma(\epsilon)\int_{0}^{\infty}dz~\frac{\left|z^2 -
(1+z)\alpha\right|^{\epsilon/2}}{z^2 - (1+z)\alpha}\ ,
\end{eqnarray}\noindent where $\alpha \equiv - p^2/m^2\,.$
Similar manipulations with the second term in Eq.~(\ref{diagenk7})
give
\begin{eqnarray}\label{int3}
L(p) &\equiv& \mu^{\epsilon}\int \frac{d^{4-\epsilon}
k}{(2\pi)^3}\frac{\delta[(q+k)^2 - m^2]}{[k^2 + i0][(k-p)^2 + i0]}
= - \frac{\mu^{\epsilon}}{(2\pi)^4}\iiint_{0}^{\infty}dx dy dz
\nonumber\\&\times& \int d^{4-\epsilon} k\left(e^{ix(k^2 + 2kq)} +
e^{-ix(k^2+2kq)}\right)e^{iy(p^2 - 2k(q+p) + i0)}e^{iz(-2kq + i0)} \nonumber\\
&=& \frac{i\mu^\epsilon}{(2\pi)^4}\iiint_{0}^{\infty}dx dy dz
\left(\frac{\pi}{x}\right)^{2 - \epsilon/2}e^{iyp^2}
\left\{\exp\left[- \frac{i}{x}((x-z)q - y(q + p))^2\right]
\right.\nonumber\\ &-& \left. \exp\left[\frac{i}{x}((x+z)q + y(q +
p))^2\right]\right\} = \frac{\pi^{-\epsilon/2}}{16\pi^2 m^2
}\left(\frac{\mu}{m}\right)^{\epsilon}\Gamma\left(1 +
\frac{\epsilon}{2}\right)\nonumber\\ &\times&\iint_{0}^{\infty} dy
dz\left\{\frac{e^{i\pi\epsilon/4}}{[(y + z + 1)^2 + yz\alpha]^{1 +
\epsilon/2}} + \frac{e^{-i\pi\epsilon/4}}{[(y + z - 1)^2 +
yz\alpha]^{1 + \epsilon/2}} \right\}\,. \nonumber
\end{eqnarray}\noindent Changing the integration variables $y = ut,$
$z = (1 - u)t$ yields
\begin{eqnarray}\label{int4}
L(p) &=& \frac{\pi^{-\epsilon/2}}{16\pi^2 m^2
}\left(\frac{\mu}{m}\right)^{\epsilon}\Gamma\left(1 +
\frac{\epsilon}{2}\right)\nonumber\\ &\times&\int_{0}^{\infty} dt
\int_{0}^{1}du~t\left\{\frac{e^{i\pi\epsilon/4}}{[(t + 1)^2 + u(1
- u)t^2\alpha]^{1 + \epsilon/2}} + \frac{e^{-i\pi\epsilon/4}}{[(t
- 1)^2 + u(1 - u)t^2\alpha]^{1 + \epsilon/2}} \right\}\,.
\nonumber
\end{eqnarray}\noindent On the other hand, $L(p)$ must be real
because the poles of the functions $D_{++}(k),$ $D_{++}(k - p)$
actually do not contribute. Hence,
\begin{eqnarray}\label{int5}
L(p) &=& \frac{\pi^{-\epsilon/2}}{8\pi^2 m^2
}\left(\frac{\mu}{m}\right)^{\epsilon}\cos\left(\frac{\pi\epsilon}{4}\right)
\Gamma\left(1 + \frac{\epsilon}{2}\right)\int_{0}^{\infty} dt
\int_{0}^{1}\frac{t~d u}{[(t + 1)^2 + u(1 - u)t^2\alpha]^{1 +
\epsilon/2}}\ . \nonumber
\end{eqnarray}\noindent

Let us turn to investigation of singularities of the expressions
obtained when $\epsilon \to 0.$ Evidently, both $K(p)$ and $L(p)$
contain single poles
\begin{eqnarray}\label{div1}
K_{\rm div}(p) &=& \frac{1}{8\pi^2
m^2\epsilon}~\dashint_{0}^{\infty}\frac{dz}{z^2 - (1+z)\alpha} = -
\frac{1}{8\pi^2 m^2\epsilon}\frac{1/\alpha}{\sqrt{1 +
4/\alpha}}\ln\frac{\sqrt{1 + 4/\alpha} + 1}{\sqrt{1 + 4/\alpha} -
1}\ , \nonumber\\ \label{div2} L_{\rm div}(p) &=& \frac{1}{8\pi^2
m^2\epsilon}~ \int_{0}^{1}\frac{du}{[1 + u(1 - u)\alpha]} =
\frac{1}{4\pi^2 m^2\epsilon}\frac{1/\alpha}{\sqrt{1 +
4/\alpha}}\ln\frac{\sqrt{1 + 4/\alpha} + 1}{\sqrt{1 + 4/\alpha} -
1}\ .
\end{eqnarray}\noindent Note that $L_{\rm div} = - 2 K_{\rm div}.$
Upon substitution into Eq.~(\ref{diagenk7}) the pole terms cancel
each other. Thus, $\tilde{C}_{\mu\nu}(p,q)$ turns out to be finite
in the limit $\epsilon \to 0.$ Taking into account also that
divergences of the remaining two diagrams (g), (h) in
Fig.~\ref{fig1} are independent of the momentum
transfer,\footnote{The corresponding integrals do not involve
dimensional parameters other than $p,$ and therefore are functions
of $p^2$ only. Hence, on dimensional grounds, both diagrams are
proportional to
$$\left(\frac{c_1}{\epsilon} +
c_2\right)\left(\frac{\mu^2}{-p^2}\right)^{\epsilon/2} =
\frac{c_1}{\epsilon} + c_2 +
\frac{c_1}{2}\ln\left(\frac{\mu^2}{-p^2}\right) + O(\epsilon),$$
where $c_{1,2}$ are some finite constants.} we conclude that the
singular part of the correlation function is completely local.

It is worth of mentioning that although the quantity $K_{\rm
div}(p)$ is non-polynomial with respect to $p^2,$ signifying
non-locality of the corresponding contribution to
$C_{\mu\nu}(x,x'),$ it is analytic at $p = 0,$ which implies that
its Fourier transform is local to any finite order of the
long-range expansion. Indeed, expansion of $K_{\rm div}(p)$ around
$\alpha = 0$ reads
\begin{eqnarray}\label{div11}
K_{\rm div}(p) &=& - \frac{1}{16\pi^2 m^2\epsilon}\left(1 -
\frac{\alpha}{6} + \cdots + \frac{(-1)^n(n!)^2}{(2n + 1)!}\alpha^n
+ \cdots\right)\,.
\end{eqnarray}\noindent
Contribution of such term to $C_{\mu\nu}(x,x')$ is proportional to
\begin{eqnarray}
\int\frac{d^3 \bm{p}}{(2\pi)^3}e^{i\bm{p}(\bm{x'} - \bm{x}_0
)}\left(1 - \frac{\alpha}{6} + \cdots\right) =
\delta^{(3)}(\bm{x'} - \bm{x}_0) +
\frac{1}{6m^2}\triangle\delta^{(3)}(\bm{x'} - \bm{x}_0) +
\cdots\,. \nonumber
\end{eqnarray}
\noindent The delta function arose here because we neglected
spacial spreading of the wave packet. Otherwise, we would have
obtained
\begin{eqnarray}
\iint\frac{d^3 \bm{q}}{(2\pi)^3}\frac{d^3
\bm{p}}{(2\pi)^3}b^*(\bm{q})b(\bm{q} + \bm{p}) e^{i\bm{p}(\bm{x'}
- \bm{x}_0 )}\left(1 - \frac{\alpha}{6} + \cdots\right) = 0 \quad
{\rm for} \quad \bm{x'} - \bm{x}_0 \notin W\,, \nonumber
\end{eqnarray}
\noindent as a consequence of the condition (\ref{packet}). In
particular, applying this to the correlation function, we see that
its divergent part does not contribute outside of $W.$ Thus, the
two-point correlation function has a well defined coincidence
limit everywhere except the region of particle localization.

Turning to calculation of the finite part of $K(p),$ we subtract
the divergence from the right hand side of Eq.~(\ref{int1}), and
set $\epsilon = 0$ afterwards:
\begin{eqnarray}\label{int1f}&&
K_{\rm fin}(p)\equiv \lim\limits_{\epsilon\to 0 }\bigl[K(p) -
K_{\rm div}(p)\bigr] \nonumber\\&& = \frac{1}{16\pi^2
m^2}\Biggl\{- \left[i\pi + \ln\pi + \gamma + 2\ln\left(\frac{-
p^2}{\mu m}\right) \right] \dashint_{0}^{\infty}\frac{dz}{z^2 -
(1+z)\alpha} \nonumber\\&&+
~\dashint_{0}^{\infty}dz~\frac{\ln\left|z^2 -
(1+z)\alpha\right|}{z^2 - (1+z)\alpha}\Biggr\}\ ,
\end{eqnarray}\noindent where $\gamma$ is the Euler constant.
The leading contribution is contained in the last integral.
Extracting it with the help of Eq.~(\ref{root}) of the Appendix,
we find
\begin{eqnarray}\label{int1l}
K^{(0)}_{\rm fin}(p) = \frac{1}{64m\sqrt{-p^2}}\,.
\end{eqnarray}\noindent

As to $L(p),$ it does not contain the root singularity. Indeed,
\begin{eqnarray}
L(0) = \frac{\pi^{-\epsilon/2}}{8\pi^2 m^2
}\left(\frac{\mu}{m}\right)^{\epsilon}\cos\left(\frac{\pi\epsilon}{4}\right)
\frac{\Gamma\left(1 + \epsilon/2\right)}{\epsilon(1 +
\epsilon)}\,,\nonumber
\end{eqnarray}\noindent and therefore $L_{\rm fin}(p)\equiv
\lim\limits_{\epsilon\to 0 }[L(p) - L_{\rm div}(p)]$ is finite at
$p=0.$ It is not difficult to verify that $L_{\rm fin}(p)$ is in
fact analytic at $p = 0.$ Thus, substituting Eq.~(\ref{int1l})
into Eqs.~(\ref{diagenk7}), (\ref{diagenk3}), and then into
Eq.~(\ref{c0i0}), and using the formula
\begin{eqnarray}
\int \frac{d^3\bm{p}}{(2\pi)^3}\frac{e^{i\bm{p x}}}{|\bm{p}|} &=&
\frac{1}{2\pi^2 \bm{x}^2}\,, \nonumber
\end{eqnarray}
\noindent  we finally arrive at the following expression for the
leading long-range contribution to the correlation function
\begin{eqnarray}\label{final}
\EuScript{C}_{00}^{(0)}(x) = \frac{e^2}{32\pi^2r^2}\ , \qquad r =
|\bm{x} - \bm{x}_0|\,,
\end{eqnarray}\noindent all other components of
$\EuScript{C}^{(0)}_{\mu\nu}$ vanishing. This result coincides
with that obtained by the author in the framework of the S-matrix
approach \cite{kazakov3}. The root mean square fluctuation of the
Coulomb potential turns out to be
\begin{eqnarray}\label{finalrms}
\sqrt{\left\langle A^2_0(r) \right\rangle} =
\frac{e}{\sqrt{32}\,\pi r}\,.
\end{eqnarray}\noindent Note also that the relative value of the
fluctuation is $1/\sqrt{2}.$ It is interesting to compare this
value with that obtained for vacuum fluctuations. As was shown in
Ref.~\cite{zerbini}, the latter is equal to $\sqrt{2}$ (this is
the square root of the relative variance $\Delta^2_r$ used in
Ref.~\cite{zerbini}).

We can now ask for conditions to be imposed on a system in order
to allow classical consideration of its electromagnetic
interactions. Such a condition can easily be found out by
examining dependence of the $\hbar^0$ contribution on the number
of field producing particles. Let us consider a body with total
electric charge $Q$ and mass $M,$ consisting of a large number $N
= Q/q$ of elementary particles with charge $q$ and mass $m,$
assumed identical for simplicity. Then it is readily seen that the
diagrams (b)--(f) in Fig.~\ref{fig1} are proportional to $N\cdot
q^2 = Q^2/N\,$ because they have only two external matter lines.
We conclude that the $\hbar^0$ contribution to the correlation
function, given for an elementary particle by Eq.~(\ref{final}),
turns into zero in the macroscopic limit $N\to \infty.$ At the
same time, the next term in the long-range expansion of the
correlation function has the form
$$\EuScript{C}^{(1)}\sim \frac{e^2}{m}\frac{\hbar}{cr^3}\,.$$ Hence, for a
fixed mass $M$ of the multi-particle body, it is independent of
$N,$
$$\EuScript{C}^{(1)}\sim N\frac{(Q/N)^2}{(M/N)}\frac{\hbar}{cr^3} =
\frac{Q^2}{M}\frac{\hbar}{cr^3}\,,$$ thus recovering the classical
estimate $\mathfrak{S} = O(\sqrt{\hbar})$ in the macroscopic
limit.

\section{Discussion and conclusions}\label{conclud}

The main result of the present work is the expression
(\ref{finalrms}) describing quantum fluctuations of the
electrostatic potential of a charged particle. It represents the
leading contribution in the long-range expansion, and is valid at
distances much larger than the Compton length of the particle.
Remarkably, the fluctuation turns out be zero order in the Planck
constant. This result is obtained by evaluating two-point
correlation function of the electromagnetic 4-potential in the
coincidence limit. We have shown that despite non-locality of
divergences arising in various diagrams in this limit, the total
divergent part of the in-in matrix element is local, and hence
does not contribute outside the region where the particle is
localized.

Perhaps, it is worth to stress once more that the issue of
locality of divergences is only a technical aspect of our
considerations. This locality does make the structure of the
long-range expansion transparent and comparatively simple.
However, even if the divergence were nonlocal this would not
present a principal difficulty. A physically sensible definition
of an observable quantity always includes averaging over a finite
spacetime domain, while the singularity of the two-point function,
occurring in the coincidence limit, is integrable (see
Sec.~\ref{lr}). The only problem with the nonlocal divergence
would be impossibility to take the limit of vanishing size of the
spacetime domain [Cf. Eq.~(\ref{c0i0})].

Turning back to the problem of measurability of electromagnetic
field, which served as the starting point of our investigation, we
may conclude that the proof of principal realizability of
measurements, given in Refs.~\cite{bohr1,bohr2}, is incomplete,
since it relies heavily on the $O(\sqrt{\hbar})$ estimate for the
electromagnetic field fluctuations, which is inapplicable to the
fields of elementary particles. The latter case therefore requires
further investigation.

As we saw in the end of Sec.~\ref{calcul}, the $\hbar^0$
contribution to the correlation function of the electromagnetic
field of a macroscopic body is suppressed by the factor $1/N.$
This is in agreement with the necessary condition for neglecting
the $\hbar^0$ contribution, formulated in Ref.~\cite{kazakov2} in
connection with the loop corrections to the post-Newtonian
expansion in general relativity. In the present context, it
coincides with the well-known criterion for neglecting quantum
fluctuations \cite{tomboulis,callan,zerbini}.

\acknowledgments{I thank Drs. A.~Baurov, P.~Pronin, and
K.~Stepanyantz (Moscow State University) for useful discussions.}

\begin{appendix}

\section{}

\begin{figure}
\includegraphics{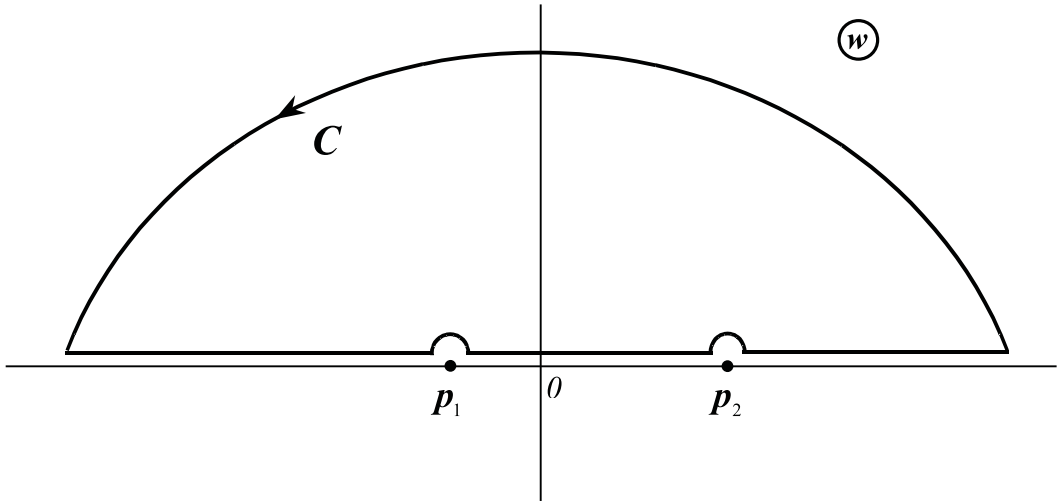}\caption{Contour of integration in
Eq.~(\ref{cint}).}\label{fig2}
\end{figure}

In the course of separation of the finite part in the right hand
side of Eq.~(\ref{int1}) we encountered the integral
\begin{eqnarray}
A \equiv \dashint_{0}^{\infty}dz~\frac{\ln\left|z^2 - (1 + z
)\alpha\right|}{z^2 - (1 + z )\alpha} &=&
\dashint_{0}^{\infty}dz~\frac{\ln\left|(z - p_1)(z -
p_2)\right|}{(z - p_1)(z - p_2)}\,, \nonumber\\ p_{1,2} &=&
\frac{\alpha \pm\sqrt{\alpha^2 + 4\alpha}}{2}\,. \nonumber
\end{eqnarray}
\noindent To evaluate this integral, let us consider the integral
\begin{eqnarray}\label{cint}
\bar{A} = \int_{C}d w~\frac{\ln[(w - p_1)(w - p_2)]}{(w - p_1)(w -
p_2)}\ ,
\end{eqnarray}
\noindent taken over the contour $C$ in the complex $w$ plane,
shown in Fig.~\ref{fig2}. $\bar{A}$ is zero identically. On the
other hand, one has for sufficiently small positive $\sigma$
\begin{eqnarray}&&
\bar{A} = \int_{-\infty}^{p_1 - \sigma}dz~\frac{\ln[(z - p_1)(z -
p_2)]}{(z - p_1)(z - p_2)} - i\pi \frac{\ln(p_2 - p_1)}{p_1 - p_2}
- \frac{i\pi\ln\sigma + \pi^2/2}{p_1 - p_2} \nonumber\\&& +
\int_{p_1 + \sigma}^{p_2 - \sigma}dz~\frac{\ln[(z - p_1)(p_2 - z)]
- i\pi}{(z - p_1)(z - p_2)} - i\pi \frac{\ln(p_2 - p_1) -
i\pi}{p_2 - p_1} - \frac{i\pi\ln\sigma + \pi^2/2}{p_2 - p_1}
\nonumber\\&& + \int_{p_2 + \sigma}^{\infty}dz~\frac{\ln[(z -
p_1)(z - p_2)] - 2i\pi}{(z - p_1)(z - p_2)}\,. \nonumber
\end{eqnarray}
\noindent Changing the integration variable $z \to (p_1 + p_2) -
z$ in the first integral, and rearranging yields in the limit
$\sigma \to 0$
\begin{eqnarray}\label{eroot}
A = \frac{1}{2}\int_{0}^{p_1 + p_2} dz~\frac{\ln[(p_2 - z)(z - p_1
)]}{(z - p_1)(z - p_2)} + \frac{\pi^2/2}{p_2 - p_1}\,.
\end{eqnarray}
\noindent The root singularity is contained in the second term,
because
$$\int_{0}^{p_1 + p_2} dz~\frac{\ln[(p_2 - z)(z - p_1 )]}{(z -
p_1)(z - p_2)} \to - \ln\alpha \quad {\rm for} \quad \alpha\to
0.$$ Thus,
\begin{eqnarray}\label{root}
A^{\rm root} = \frac{\pi^2}{4\sqrt{\alpha}}\,.
\end{eqnarray}
\noindent It is interesting to note also that the part containing
the root singularity [the second term in Eq.~(\ref{eroot})]
exactly coincides with that found using the S-matrix (Cf. Eq.~(B2)
in Ref.~\cite{kazakov3}).

\end{appendix}

\pagebreak

\pagebreak

\end{document}